\documentclass[%
 aip,
 amsmath,amssymb,
reprint,%
]{revtex4-1}
\usepackage{graphicx}
\usepackage{dcolumn}
\usepackage{bm}
\usepackage{lipsum}
\usepackage[utf8]{inputenc}
\usepackage[T1]{fontenc}
\usepackage{mathptmx}
\usepackage{etoolbox}
\usepackage{tabularx}
\usepackage{xcolor}
\graphicspath{{./figures/}}

\makeatletter
\def\@email#1#2{%
 \endgroup
 \patchcmd{\titleblock@produce}
  {\frontmatter@RRAPformat}
  {\frontmatter@RRAPformat{\produce@RRAP{*#1\href{mailto:#2}{#2}}}\frontmatter@RRAPformat}
  {}{}
}%
\makeatother
\begin{document}

\preprint{AIP/123-QED}

\title{Hybrid superinductance with Al/InAs}

\author{Junseok Oh}
\affiliation{The James Franck Institute, University of Chicago, Chicago, Illinois 60637, USA}
\author{Ido Levy}
 \affiliation{Center for Quantum Phenomena, Department of Physics, New York University, New York 10003, USA}
\author{Tyler Cowan}
 \affiliation{Center for Quantum Phenomena, Department of Physics, New York University, New York 10003, USA}
\author{Jacob Issokson}
 \affiliation{Center for Quantum Phenomena, Department of Physics, New York University, New York 10003, USA}
\author{Archana Kamal}
\affiliation{Department of Physics and Astronomy, Northwestern University, Evanston, Illinois 60208, USA}
\author{Javad Shabani}
\affiliation{Center for Quantum Phenomena, Department of Physics, New York University, New York 10003, USA}
\author{Andrew P. Higginbotham}
\affiliation{The James Franck Institute, University of Chicago, Chicago, Illinois 60637, USA}
\affiliation{The Department of Physics, University of Chicago, Chicago, Illinois 60637, USA}
\date{\today}

\begin{abstract}
We report microwave spectroscopy of Josephson junctions chains made from an epitaxial Al/InAs heterostructure.
The chains exhibit superinductance, with characteristic wave impedance exceeding $R_{Q} = \hbar/(2e)^{2}$.
The planar nature of the junctions results in a large plasma frequency, with no measurable deviations from ideal dispersion up to $12~\mathrm{GHz}$.
Internal quality factors decrease sharply with frequency, which we describe with a simple loss model.
The possibility of a loss mechanism intrinsic to the superconductor-semiconductor junction is considered.
\end{abstract}

\maketitle

\section{Introduction}
Superconducting circuits are currently a topic of interest, driven by the growth of superconducting quantum computing \cite{blais2021circuit}.
High-impedance circuits, most notably the fluxonium qubit \cite{manucharyan2009fluxonium}, have appealing properties such as high coherence \cite{nguyen2019high}, high-fidelity gates \cite{bao2022fluxonium,ding2023high}, and routes to noise protection \cite{gyenis2021experimental,ardati2024using,mencia2024integer}.
The enabling circuit element for high-impedance qubits is a superinductor -- a device whose characteristic impedance $Z$ exceeds the resistance quantum $R_Q~=~\hbar/4e^2~\approx~1.027 $ k$\Omega$ \cite{masluk2012microwave}.
Creating ideal superinductance remains challenging because of stringent and often conflicting requirements. 
Take superinductance realized by a chain of Josephson junctions as an example.
In a tunnel junction, avoiding phase slips requires a small charging energy $E_C \ll E_J$ where $E_C$ is the junction charging energy and $E_J$ is the Josephson energy.
However small $E_C$ comes at a cost, since it leads to small plasma frequency $\omega_p \sim \sqrt{E_C E_J}$, reducing operating frequencies.
Thus, there are conflicting constraints on $E_C$ that must be navigated.

The challenge of creating ideal superinductance has recently motivated exploration of alternative approaches \cite{maleeva2018circuit,kuzmin2019quantum,mukhopadhyay2023superconductivity,charpentier2025first,charpentier2025universal}.
Superconductor-semiconductor hybrids are an interesting system for consideration.
The ability to make gate-tunable, field-resilient, and high-transparency Josephson junctions in this system has already yielded a series of exciting developments in the demonstration of gate-tunable transmon qubits \cite{larsen2015semiconductor,luthi2018evolution,casparis2018superconducting}, Andreev qubits \cite{hays2018direct,tosi2019spin}, and amplifiers \cite{phan2023gate,hao2024kerr}.
Recently the first fluxonium hybrid qubit was demonstrated using an Al/InAs array as the inductive element \cite{strickland2025gatemonium}, posing the new problem of hybrid superinductance.

Hybrid superinductance is radically different from tunnel-junction superinductance.
Due to planar geometry, hybrid junctions have a large charging energy; $E_C>E_J$ is nearly unavoidable.
While yielding the benefit of high plasma frequency, the presence of a large charging energy immediately raises the potential drawback of quantum phase slips.
On the other hand, in hybrid junctions phase slips may be suppressed due to high transparency \cite{bargerbos2020observation,kringhoj2020supressed}, which occurs without fine tuning in the diffusive limit \cite{vanevic2012quantum}.
Does hybrid superinductance confer the ability to break the superinductance ``barrier'' $E_C/E_J > 1$, yielding high plasma frequency?
Or, rather, is performance degraded by phase slips?
The above conundrum motivated our study.

We fabricate chains of Josephson junctions from an Al/InAs 2D electron gas and investigate their microwave response.
In order to obtain a sufficiently large inductance without stray capacitance from electrostatic gates, we explore arrays in the long-junction limit\cite{mayer2019superconducting}.
Comparing devices with different junction sizes, we reach the superinductance regime $Z>R_Q$, and, unlike conventional Al/AlOx devices, find no resolvable limitations from the single-junction plasma frequencies.
The observed dispersion is well described by finite-element microwave simulations.
Quality factors decrease sharply with increasing mode frequency, which is the opposite of the behavior expected from phase slips.
Rather, we find that loss is well described by an effective resistance shunting the junctions.
Although the shunt's microscopic origin is not certain, we suggest that it is related to our use of junctions in the diffusive limit, and discuss mitigation strategies that would make our superinductance approach practical for fluxonium qubits.
At current quality levels, our devices are promising for readout of spin \cite{reilly2007fast,petersson2010charge,crippa2019gate,bartee2025spin} and parity qubits \cite{aghaee2025interferometric,vanloo2025single,aghaee2025distinct} at modest frequencies ($<1~\mathrm{GHz}$), potentially outperforming dominant coil inductors \cite{hornibrook2014frequency} by offering higher quality factor, reduced capacitance, and compact form factor.

\section{Experiment and Results}
\begin{figure}
\includegraphics{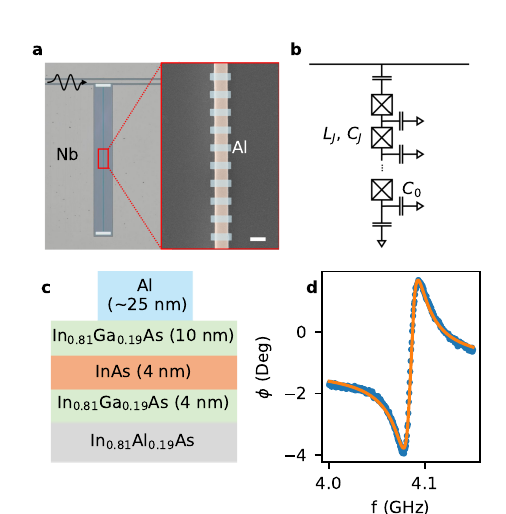}
\caption{(a) Left: Optical image of a chain of 800 Josephson junctions is capacitively coupled to a coplanar waveguide (CPW) in the hanger geometry.
Right: the close-up image taken with a scanning electron microscope, Al islands (blue) above the InAs 2D electron gas (orange) are separated by the junction length $l_J$, forming Josephson junction chains.
The unit cell size is $d=1.2~\mathrm{\mu m}$ for all devices. Scale bar on the bottom right is 1 $\mu$m. (b) Lumped element model of a Josephson junction chain.
(c) Top layers of the material stack. InGaAs/InAs/InGaAs layers form 2D electron gas.
The two capacitors at each end of the chain represents coupling capacitors. (d) $S_{21}$ phase data (points) and the fit (solid line) of the mode 3 of the 700 nm Device  1 at the base temperature.}
\end{figure}

The devices are on-chip resonators capacitively coupled to a coplanar waveguide in the hanger geometry (Fig.~1(a)).
Each resonator consists of 800 planar Al/InAs Josephson junctions in series.
The material stack shown in Fig.~1(c) is prepared by epitaxial growth.
The mobility and the carrier density of the film was determined with Hall measurement to be $(1.5\pm0.2)\times10^{4} ~\mathrm{cm^{2}/Vs}$ and $9.76\times10^{11} ~\mathrm{cm^{-2}}$.
We use electron-beam lithography to pattern and selective etching to form Al islands, followed by similar steps to form $1~\mathrm{\mu m}$-wide pillar-like structure of semiconducting layers, commonly referred to as ``mesa.''
The Al islands of the resulting devices are separated by junction length $l_J$ and unit cell size $d = 1.2~\mathrm{\mu m}$ to form planar Josephson junctions.

Three devices are fabricated and measured: two devices with $l_J =$ 700 nm (Device 1,2) and a device with $l_J =$ 400 nm (Device 3).
The Josephson junction chain can be understood as an equivalent circuit shown in Fig.~1(b).
Each Josephson junction has a shunt capacitance $C_J$, an equivalent inductance $L_J$ given by the Josephson relation, and a parasitic capacitance $C_0$ to ground.
The experimental microwave setup is standard for circuit quantum electrodynamics experiments. 
Measurements are performed using a vector network analyzer and a dilution refrigerator with a base temperature of $23~\mathrm{mK}$.
The microwave input line is equipped with attenuators for thermalization.
On the readout line, a double-junction 4-8 GHz isolator at base temperature is followed by a cryogenic high electron mobility transistor amplifier on the 4 Kelvin stage and a room temperature amplifier.


In Device 1, measured microwave transmission shows a resonance near $4.1~\mathrm{GHz}$ (Fig.~1(d)), yielding resonator parameters $Q_{~\mathrm{tot}} = 257$, $Q_{~\mathrm{ext}} = 2830$, and $Q_{i} = 284$ when fit\cite{probst2015efficient}.
The fit  $Q_\mathrm{ext}$ is reasonably close to expectations based on finite-element simulations ($4120$), indicating that the external coupling is understood.

To identify other modes in a broader frequency range, two-tone spectroscopy was used; while measuring the $S_{21}$ of the resonance at 4 GHz with the vector network analyzer, a secondary pump tone was swept through a wide frequency range.
When the pump tone is resonant with j\textsuperscript{th} mode of the resonator, there is a cross-Kerr shift $\delta f_i$ in the i\textsuperscript{th} mode which can be detected by a vector network analyzer.
For all two-tone measurements, $\delta f_3$ was measured for Device 1 and 2, and $\delta f_1$ for Device 3.
Fig. 2(a-d) shows the measured phase shift $\delta f_3$ near $j=3$ mode of the Device 1 while the pump tone is swept through mode 1-4.
The pump power was carefully kept low to ensure a ``linear response'' regime ($\delta f_i = K_{ij} \langle n_j \rangle$), allowing the mode line-shape to be faithfully extracted.
Here, $K_{ij}$ is the cross-Kerr coefficient for modes $i$ and $j$, and $\langle n_j \rangle$ is the average number of photons in mode $j$\cite{weissl2015kerr,krupko2018kerr}.
A Lorentzian fit was applied to the two-tone data to obtain the resonance frequencies and loaded quality factors.
We first study the mode frequencies, and then the quality factors.

\begin{figure}
\includegraphics{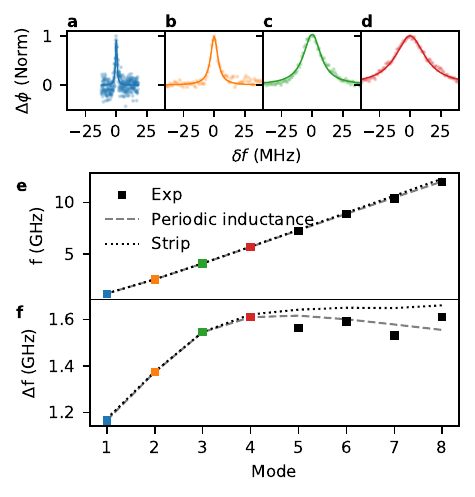}
\caption{(a-d) Mode 1 - 4 of Device 1 measured through two-tone spectroscopy.
The phase shift $\Delta \phi$ of the mode 3 was measured while a secondary pump was swept through each mode.
The pump power was kept sufficiently low to stay in linear regime.
The solid lines are fits to Lorentzian.
(e) Mode frequencies and (f) mode spacing ($\Delta f$) extracted from the fit as a function of mode number.
Up to mode 8, the dispersion is nearly linear.
The dashed grey line represents the dispersion simulated using an EM solver with 800 unit cells of alternating inductance.
The dotted black line is the simulation result for a strip with a uniform inductance.}
\end{figure}

The extracted mode frequencies and mode spacings of Device 1 are plotted in Fig. 2(e) and (f), respectively.
Fig.~2(e) shows a nearly linear dispersion relation up to mode 8.
Above mode 8, the $Q$ becomes too small to yield a reliable Lorentzian fit.
A more detailed view of the mode dispersion can be gained by examining the mode spacing, $\Delta f_{i} = f_{i}-f_{i-1}$ with $f_{0} = 0$ (Fig. 2(f)).
The mode spacing initially increases up to mode 4 and decreases above mode 4, showing a slight deviation from the linear dispersion.
The initial increase in the mode spacing can be attributed to the coupling capacitors at the end of the chain\cite{masluk2012microwave}.
To gain quantitative understanding of the mode spacing, we performed finite-element simulations of the device, representing junctions as regions of large sheet inductance and Al as regions with vanishing sheet inductance.
A junction inductance of $1.8~\mathrm{nH}$ per unit cell accurately reproduces the measured dispersion (Fig. 2(e-f)).
Thus, the slight curvature in the dispersion does not reflect a Josephson plasma frequency -- it is a consequence of the electromagnetic environment and device geometry.
Combined with the measured speed of light in the chain (mode spacing $\Delta f\approx f_4 -f_3 \approx 1.5 ~\mathrm{GHz}$), we find a chain impedance of $4.7~\mathrm{kOhm}$, indicating that we have successfully realized an Al/InAs hybrid superinductor.
Through the same calculation, $C_0$ is estimated to be $\approx 80~\mathrm{aF}$.
As shown in Fig. 2(f), a distributed strip inductance shows less dispersion than a periodic inductor, and does not match the experiment as well.
We conclude that, while there is no indication of the shunt capacitance affecting dispersion, periodic inductance does play a role.

The same procedure was used to extract dispersion, mode spacing, and inductance for other measured devices (Fig.~3).
For the second 700 nm device (Device 2), we find a similar dispersion, Josephson inductance (2.14 nH), and characteristic impedance (5.17~$\mathrm{k\Omega}$), compared to Device 1.
In contrast, the device with 400 nm junction length exhibits larger mode spacing, lower Josephson inductance (0.11 nH) and lower wave impedance ($1.07~\mathrm{k\Omega}$).
The drastic difference in inductance between 400 nm and 700 nm junction size is expected based on their relative lengths \cite{mayer2019superconducting}.
Reproducibility between two devices and control over the dispersion indicates that our devices are well controlled.
We notice the 400 nm device shows a larger deviation from the simulation compared to the 700 nm devices, which we attribute to microwave problems specific to that cooldown.

For all devices, the finite element simulation of periodic inductance captures the observed dispersion without any additional junction shunting capacitance.
This is consistent with the fact that the $C_J$ is vanishingly small for planar junction.
In contrast, standard Al/AlOx superinductance typically has large dispersion \cite{masluk2012microwave,manset2025hyperinductance}, dominated by the Josephson shunt capacitance.
The negligible shunt capacitance of planar Al/InAs junctions may therefore open the door to higher-frequency Josephson superinductance.

It is interesting that small $C_J$ makes a nominal criterion for superinductance~\cite{manucharyan2009fluxonium}, $N > \sqrt{C_J/C_0}$, difficult to achieve.
The criterion is an assertion that array self-resonant frequencies should be larger than the single-junction plasma frequency.
For planar junctions, where the junction plasma frequency is immeasurably large, the criteria is no longer applicable.
Rather, one need only ensure that array self-resonances are outside of the band of interest.

\begin{figure}
\includegraphics{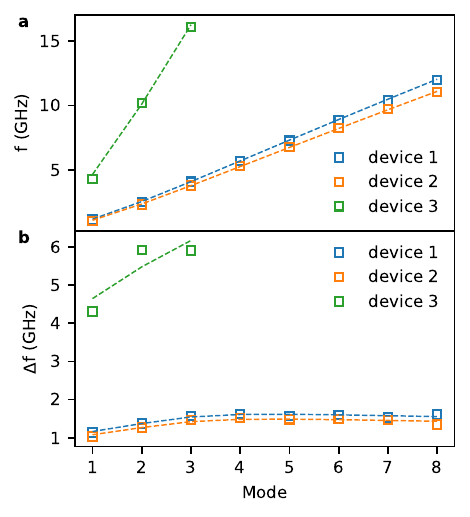}
\caption{(a) The mode frequencies and (b) the mode spacing for all measured devices.
EM simulation results with periodic inductance are shown in dashed lines.
}
\end{figure}

\begin{figure}
\includegraphics{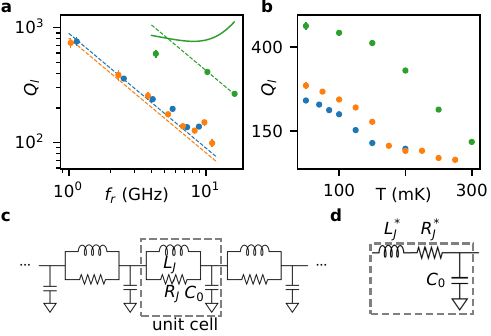}
\caption{(a) $Q_{l}$ as a function of frequency for all devices in log-log scale.
Dashed lines represent fits to Eq.~\ref{eq:qi} assuming $Q_c \gg Q_i$ such that $Q_l \approx Q_i$.
For device 3 the simulated $Q_c$ (solid line) approaches $Q_i$ for the lowest-frequency point, which is therefore excluded from the fit.
For devices 1 and 2, $Q_c$ is much larger than $Q_l$ for all points.
Each color correspond to a device as in Fig. 3. (b) Temperature dependence of the $Q_l$ for a mode near 4 GHz (mode 3 for Device 1,2 and mode 1 for device 3).
Temperatures $\geq 50~\mathrm{mK}$ shown, for thermalization reasons.
(c) Lumped circuit model of a resistively shunted Josephson junction. (d) Equivalent series circuit model of a unit cell.
}
\end{figure}

The $Q$ values of our Al/InAs superinductors are orders of magnitude smaller than Al/AlOx superinductors \cite{masluk2012microwave}.
$Q_l$ decreases dramatically with mode frequency, decreasing by an order of magnitude between $1-10~\mathrm{GHz}$, and obeys an approximate inverse frequency relationship for Devices 1 and 2 (Fig. 4(a)).
The observed quality factors are too small in magnitude to be attributed to the external coupling.
Furthermore, the frequency dependence of the observed quality factor is opposite from that expected for external coupling.
We therefore conclude that the quality factor is limited by internal loss.
The order-of-magnitude change that we observe with increasing frequency is too strong to be associated with dielectric loss \cite{gau_physics_2008,kaiser2010measurement}.
We therefore propose a phenomenological loss model of a resistively shunted Josephson junction that naturally explains the frequency dependence.
Qualitatively, the frequency-dependent loss arises from enhanced participation of the shunt resistance as frequency increases (a similar effect was reported in Phan et al\cite{phan2023gate}).

The phenomenological loss model includes the Josephson inductance $L_J$, shunt resistance $R_J$, and a parasitic capacitance $C_0$ to the ground.
The parallel circuits' impedance can be re-expressed as arising from a series equivalent resistance $R_J^{*}$ and series equivalent inductance $L_J^{*}$ 
\begin{equation}
  \begin{split}
R_J^{*} = \frac{R_J \omega^2 L_J^2}{R_J^2 + \omega^2 L_J^2}\\
L_J^{*} = \frac{L_J R_J^2}{R_J^2 + \omega^2 L_J^2}.
  \end{split}
\end{equation}
In the weak-shunt limit ($R_J > \omega L_J$), relevant here, the shunt resistance is transformed $R_J^{*} \approx \frac{\omega^2 L_J^2}{R_J}$, whereas $L_J^* \approx L_J$.
The series model is equivalent to a lossy transmission line of length $l = Nd$ with $N$ unit cells, resistance per length $R_{l}^{*} = R_J^{*}/d$, and inductance per length $L_{l}^{*} = L_J^{*}/d$.
The internal quality factor of the n\textsuperscript{th} mode with frequency $\omega_{n} = n\omega_{0}$ is then given by the known formula
\begin{equation}
Q_{i,n} = \frac{n \pi}{2\alpha^* l},
\end{equation}
where the attenuation constant is $\alpha^* = R_l^{*}/(2Z)$~\cite{Pozar}.
In terms of the parallel circuit elements, the internal quality factor $Q_i$ as a function of $f_r = \omega_n/(2 \pi)$ becomes
\begin{equation}
  \label{eq:qi}
Q_{i} = \frac{1}{f_r}\frac{Z R_J}{2 N \omega_0 L_J^2}.
\end{equation}

Equation~\ref{eq:qi} predicts that $Q_i$ scales as $1/f_r$, as observed in the experiment (most clearly for Devices 1 and 2).
Fitting to the data in Fig.~4a along with $Z$ and $L$ values extracted from the finite-element simulations, we extract the device resistance per junction $R_J$ to be $10~\mathrm{k\Omega}$, $11~\mathrm{k\Omega}$, and $3~\mathrm{k\Omega}$ for Device 1, 2, and 3.
Interestingly, $R_J$ is smaller for Device 3 which has a shorter junction length (400 nm compared with 700 nm), as expected for an intrinsic loss mechanism related to diffusive junctions.
Qualitatively, the loss mechanism is reflected in the fact that Device 3 has much higher $Q$ at similar frequencies.

To further explore the loss mechanism in our devices, we studied the temperature dependence of the microwave quality factor for modes of comparable frequency (Fig.~4(b)).
Quality factor decreases strongly with temperature, with the longer junctions (Devices 1,2) showing strong temperature dependence below 100~mK.
The strong temperature dependence is also suggestive of an intrinsic loss mechanism associated with our long junctions.

Before concluding, we note that Device 3 does not follow the $Q_{l} \propto 1/f_r$ relation as well as 700 nm devices (Device 1,2) do.
A possible reason for the discrepancy is that the for the first mode of Device 3 $Q_l$ approaches $Q_c$ (Fig~4(a)), whereas for all other modes and devices $Q_l \ll Q_c$.
Recall also that the agreement between the EM simulation result and the observed dispersion is worse in Device 3 compared to Device 1 and 2 due to a cooldown-specific microwave problem.
$Q_l$ may also have been affected.

\section{Conclusion}
In conclusion, we fabricated Al/InAs superinductors whose characteristic impedance exceeds the resistance quantum.
We find linear mode dispersion with no discernible limitations from the single-junction plasma frequency.
The observed dispersion matches a model including the periodic inductance of the array and the electromagnetic environment.
We further propose resistively shunted junction model that can explain the inverse relation between $Q_{i}$ and the mode number.

While our experiments do not conclusively determine the microscopic origin of the shunt resistance, we argue that a loss mechanism intrinsic to the junction is plausible.
Indeed, our observed quality factors are substantially lower than Al/InAs 2DEG devices with lower junction participation ratios\cite{casparis2018superconducting,phan2022detecting,phan2023gate,strickland2024superconducting}, pointing to a loss mechanism associated with the junctions.
In considering loss mechanisms intrinsic to the junctions, it is important that our devices are in the long-junction limit, $l_J > \xi$ where $\xi \sim 200~\mathrm{nm}$ is the induced coherence length.
We then expect that the junction $I_{c} R_{N}$ product is set by the Thouless energy $E_{Th}$ which scales inversely with the square of the junction length.
Here, $I_c$ and $R_N$ are the critical current and the normal state resistance of a junction, respectively.
An estimate of our Thouless energy implies an $R_N$ order of 0.1 - 1.5 kOhm, compatible with expected long-junction values \cite{mayer2019superconducting}.
A loss mechanism associated with diffusive junctions is also empirically consistent with the observation that arrays with shorter junctions (Device 3) with proportionately higher critical current and Thouless energy, exhibit higher $Q_{i}$.
Additionally, we have found that junction loss increases dramatically with temperature, more so for the longer junctions than for the shorter junctions (Fig.~4(b)), consistent with a low Thouless energy $E_{Th} \lesssim k_B T$.
Thus, we suspect that the microscopic origin of the resistive shunt is our junction geometry.

To mitigate losses, short junctions could be used as an alternative, provided either the transverse junction dimension is reduced or electrostatic gates are incorporated to maintain sufficiently large Josephson inductance.
If the parallel loss channel is reduced, hybrid superinductance would become practically appealing, especially where large plasma frequency or higher operating temperature~\cite{telkamp2025development} is required.
At current quality levels, our devices may be competitive for readout of spin \cite{reilly2007fast,petersson2010charge,crippa2019gate,bartee2025spin} and parity qubits \cite{aghaee2025interferometric,vanloo2025single,aghaee2025distinct} at modest frequency $<1~\mathrm{GHz}$, potentially outperforming dominant coil inductors \cite{hornibrook2014frequency} by offering higher quality factor, reduced capacitance, and compact form factor.

\section{Acknowledgement}

This work was funded by the Defense Advanced Research Projects Agency (DARPA) Synthetic Quantum Nanostructures (SynQuaNon) program under Grant Agreement No. HR00112420343 and the NOMIS foundation.
Initial device characterizations was performed at the Center for Nanoscale Materials, a U.S. Department of Energy Office of Science User Facility, was supported by the U.S. DOE, Office of Basic Energy Sciences, under Contract No. DE-AC02-06CH11357.
This work made use of the shared facilities at the University of Chicago Materials Research Science and Engineering Center, supported by National Science Foundation under award number DMR-2011854. 

%

\end{document}